\documentclass{article}
\usepackage{spconf,amsmath,graphicx,hyperref}
\usepackage{booktabs}

\title{MNV-17: A High-Quality Performative Mandarin Dataset for Nonverbal Vocalization Recognition in Speech}

\name{\begin{tabular}{c}
      Jialong Mai$^{1,\dagger}$, Jinxin Ji$^{2,3,\dagger}$, Xiaofen Xing$^{1,*}$, Chen Yang$^{4}$ \\ 
      Weidong Chen$^{5}$, Jingyuan Xing$^{1}$, Xiangmin Xu$^{1,6}$
      \end{tabular}
      \thanks{$^{\dagger}$These authors contributed equally to this work.}
      \thanks{$^{*}$Corresponding author: xfxing@scut.edu.cn}
      \thanks{$^{\ddagger}$ \url{https://github.com/yongaifadian1/MNV-17}}
      }

\address{$^{1}$South China University of Technology, Guangzhou, China \\
         $^{2}$The Hong Kong Polytechnic University, Hong Kong, China \\
         $^{3}$Tongji University, Shanghai, China \\
         $^{4}$Shanghai Jiao Tong University, Shanghai, China \\
         $^{5}$The Chinese University of Hong Kong, Hong Kong, China \\
         $^{6}$Foshan University, Foshan, China}

\begin{document}
\ninept
\maketitle
\begin{abstract}
Mainstream Automatic Speech Recognition (ASR) systems excel at transcribing lexical content, but largely fail to recognize nonverbal vocalizations (NVs) embedded in speech, such as sighs, laughs, and coughs. This capability is important for a comprehensive understanding of human communication, as NVs convey crucial emotional and intentional cues. Progress in NV-aware ASR has been hindered by the lack of high-quality, well-annotated datasets. To address this gap, we introduce MNV-17, a 7.55-hour performative Mandarin speech dataset. Unlike most existing corpora that rely on model-based detection, MNV-17's performative nature ensures high-fidelity, clearly articulated NV instances. To the best of our knowledge, MNV-17 provides the most extensive set of nonverbal vocalization categories, comprising 17 distinct and well-balanced classes of common NVs. We benchmarked MNV-17 on four mainstream ASR architectures, evaluating their joint performance on semantic transcription and NV classification. The dataset and pretrained model checkpoints are publicly available$^{\ddagger}$ to facilitate future research in expressive ASR.

\end{abstract}

\section{Introduction}
\label{sec:intro}

Automatic Speech Recognition (ASR) has achieved remarkable success in transcribing the lexical content of speech \cite{bai2024seed,gao2022paraformer,radford2023robust,xu2025fireredasr}, positioning it as a fundamental technology in modern human-computer interaction \cite{lin2024advancing, huang2024audiogpt, yan2024talk}. Its powerful and accurate transcription capabilities, along with robust semantic encoders, have driven advancements in a wide array of related speech tasks, such as speech synthesis \cite{zhang2025minimax, anastassiou2024seed} and emotion recognition \cite{ma2024emobox, mai2025aa}. However, human communication involves much more than just words. Mehrabian et al. \cite{mehrabian1971silent} proposed that when communicating feelings and attitudes, only 7\% of the message is conveyed through words, with 38\% being expressed through paralinguistic cues, such as non-verbal vocalizations and tone of voice. Despite this, current mainstream ASR systems are largely insensitive to the nonverbal vocalizations (NVs) embedded within speech, such as laughter, sighs, and coughs \cite{radford2023robust, gao2022paraformer}. These NVs are critical carriers of paralinguistic information, essential for a deep understanding of intent and emotion \cite{truong2007automatic, cortes2021effects}. This difficulty in recognizing NVs presents a barrier to achieving truly comprehensive and human-like interaction between humans and machines.

The challenge of integrating NV recognition into ASR is distinct from traditional Sound Event Detection (SED). Unlike SED models, which typically identify acoustic events in broader audio streams \cite{yin2025exploring}, an NV-aware ASR system must be capable of precisely locating these events within the continuous flow of semantic speech. Training such a dual capability requires data that conventional resources cannot provide: standard ASR datasets consist of speech-transcription pairs that omit NV labels \cite{yao2021wenet}, while typical audio-caption data for SED lacks the fine-grained temporal alignment with lexical content needed \cite{drossos2020clotho}. The lack of high-quality, precisely labeled domain-specific data has thus created a bottleneck, leading to a shortage of robust ASR models with strong NV recognition capabilities.

Current NV datasets exhibit limitations. First, their categorical coverage is insufficient. Some available corpora are limited to a limited number of the most common NVs \cite{10317236, wang2025capspeech}, such as laughter or breaths, which is inadequate for modeling the wide variety of human nonverbal expressions. Additionally, while recent studies have proposed NV datasets suitable for ASR or TTS, their reliance on automated, model-based annotation can lead to unreliable labels \cite{ye2025scalable, borisov2025nonverbaltts}. These datasets also often suffer from severe class imbalance, which introduces model bias \cite{mai2025aa}. This data deficiency hinders reproducible research and the development of effective models in this area.

To address the aforementioned challenges, we developed and present the Mandarin Nonverbal Vocalizations-17 (MNV-17) dataset, a 7.55-hour collection of scripted speech designed to provide a solid foundation for training and evaluating NV-aware ASR systems. The dataset, recorded by native Mandarin speakers from various regions, contains 17 distinct and common nonverbal vocalization (NV) categories with a focus on a well-balanced class distribution to reduce model bias. By using a scripted approach where speakers were prompted to perform specific NVs, we ensure that each instance is intentional and well-articulated, overcoming the ambiguity issues common in spontaneous speech datasets. To the best of our knowledge, MNV-17 features the most extensive range of NV categories among publicly available datasets.

We make two main contributions. 

First, we introduce MNV-17, a 7.55-hour Mandarin speech dataset containing 17 balanced categories of nonverbal vocalizations, created to advance research in expressive speech understanding. 

Second, we benchmarked and evaluated the joint ASR and NV recognition capabilities of mainstream ASR architectures by fine-tuning them on MNV-17. Our experimental results indicate that Large Speech Models can effectively incorporate the NV ability without compromising their lexical transcription performance.
\section{Methodology}
This section details the creation of the MNV-17 dataset, including script preparation, the recording process, and post-processing procedures.
\subsection{Script Preparation}
A major challenge in creating nonverbal vocalization datasets is mitigating class imbalance and ensuring high-quality labels. Datasets compiled from online sources often suffer from an imbalanced distribution, where common NVs like laughter or sighs vastly outnumber less frequent ones \cite{wang2025capspeech,  ye2025scalable, borisov2025nonverbaltts, 10317236}. This imbalance can bias models to favor majority classes. Furthermore, using sound event detection models for pre-annotation introduces a quality ceiling \cite{ye2025scalable, borisov2025nonverbaltts}, as the dataset's accuracy is inherently limited by the performance of the annotation model.

To overcome these issues, we adopted a scripted and performative approach to data collection. This methodology allows for precise control over the class distribution and guarantees the intentionality and clarity of each NV instance. The process for script preparation was as follows:

\textbf{Categorization of NVs:} We conducted a comprehensive survey of existing NV-aware systems and datasets to identify a broad yet distinct set of common nonverbal vocalizations \cite{darefsky2024parakeet, tang2024salmonngenerichearingabilities, guo2025fireredttsfoundationtexttospeechframework, du2024cosyvoicescalablemultilingualzeroshot, kimiteam2025kimiaudiotechnicalreport}. This resulted in the selection of 17 NV categories, which, to our knowledge, represents the most extensive collection of NV types in a publicly available dataset for NV-aware ASR.

\textbf{Controlled NV Distribution:} To realistically model real-world scenarios where multiple NVs can occur in a single utterance, we designed the scripts to contain one, two, or three NV instances. The distribution was set to a 5:3:2 ratio, respectively. When multiple NVs were included, they were randomly combined from the 17 available categories, with repetition allowed.
\begin{table*}[t]
\centering
\caption{Comparison of NV datasets. The statistics for Capspeech pertain to its nonverbal vocalization subset. MNV-17 not only features the most extensive set of NV categories but also demonstrates superior class balance, as indicated by its low max-to-min frequency ratio.}
\label{tab:dataset_comparison}
\begin{tabular}{lcccc}
\toprule
\textbf{Dataset} & \textbf{Total Duration} & \textbf{\# NV Classes} & \textbf{Class Balance (Max/Min Ratio)} & \textbf{Annotation Method} \\
\midrule
Capspeech & 1h & 3 & - & Synthesized \& Inserted \\
NonVerbalSpeech-38K & 131h & 10 & 516 & Model-based \\
NonVerbalTTS & 17h & 10 & 36 & Model-based \\
\textbf{MNV-17 (Ours)} & \textbf{7.55h} & \textbf{17} & \textbf{2.7} & \textbf{Scripted Performance} \\
\bottomrule
\end{tabular}
\end{table*}

\textbf{Contextual Script Generation using LLMs:} To ensure the NVs were embedded in natural and contextually relevant sentences, we used a Large Language Model (LLM) \cite{comanici2025gemini25pushingfrontier} for script generation. 

We employed a two-fold LLM prompting strategy to generate sentences, ensuring both categorical balance and contextual diversity across the dataset. 
Our prompt engineering was designed to create sentences with varying themes and structures.

For single NV instances, the objective was to build a robust and balanced representation for each of the 17 categories. We generated 90 unique sentences for each NV type, prompting the LLM to insert the vocalization naturally at the beginning, middle, or end of the sentence.
This method ensures comprehensive coverage of each NV in various utterance positions.

For scripts with multiple NVs, we adopted a breadth-over-depth strategy. This involved generating a smaller batch of five sentences for a much wider range of random combinations. This strategy prioritized covering a wide variety of NV interactions with fewer examples each, thereby better reflecting the diversity and complex co-occurrence of NVs in realistic speech.

Finally, two linguists verified the naturalness, semantic clarity, and diversity of all generated sentences.

\subsection{Audio Recording}

\subsubsection{Ethics Declaration}
The recording experiment involved human volunteers. Written informed consent was obtained from all participants prior to the experiment and recording. In accordance with the PLOS consent form, participants also consented to the online publication of the open dataset. The experiment was approved by the Institutional Review Board of The Hong Kong Polytechnic University.

\subsubsection{Participants}
Forty-nine participants (25 female, 24 male; mean age = 26.0 years, SD = 2.6) were recruited in Hong Kong. All participants were native Mandarin speakers born in mainland China, representing diverse geographical backgrounds (18 provinces, 20 from northern China and 29 from southern China). They had received their pre-university education in mainland China, used Mandarin for daily communication and academic purposes before the age of 18, and continued to use Mandarin in their everyday lives. Thus, their speech can be considered representative of Mandarin usage. None of the participants were professional actors or actresses, nor did they have any performance training. This was intended to elicit natural recordings, as existing datasets for paralinguistic tasks often rely on movie or performance materials \cite{Xin_2024, livingstone2018ryerson}, which previous studies have shown may compromise naturalness and accuracy \cite{busso2016msp, jurgens2015effect}.

\subsubsection{Procedure and Design}

Recordings were conducted in a soundproof room at The Hong Kong Polytechnic University, following standard procedures for phonetic experiments \cite{ji2025acoustic}. Participants were seated in front of a screen displaying the stimulus sentences. A high-quality microphone was positioned approximately 10 cm in front of the speaker’s mouth and remained fixed throughout the session. Recordings were captured in mono at a sampling rate of 44.1 kHz using Praat \cite{boersma2011praat}, which is widely used in phonetic research.

The recording sessions were conducted by the second author, a postgraduate student with professional phonetic training. Participants were not provided with sample recordings, in order to elicit natural responses based on their own interpretations. Before the task, the experimenter gave general instructions, encouraging participants to read the sentences in a natural manner.

Each participant was randomly assigned 50 sentences, except for the final participants, who read 66 sentences. Each sentence was presented on a slide, and participants paused for 3–5 seconds between sentences while the experimenter advanced the slides. If participants made an error, they restarted from the sentence where the mistake occurred. Incorrectly produced sentences were excluded from the dataset.

\subsection{Data Post-processing}
A significant challenge arose from the recording format, where multiple sentences were captured in a single continuous audio file, separated only by a few seconds of silence. Segmenting these long recordings into discrete, sentence-level samples was a critical post-processing step.

Following common data processing practices \cite{he2024emiliaextensivemultilingualdiverse}, we initially experimented with a Voice Activity Detection (VAD) model \cite{bredin2020pyannote} to segment the audio based on silences. This approach proved problematic as the VAD model could not reliably distinguish between the intended pauses between sentences and natural intra-sentence pauses, such as those occurring at commas or during the articulation of quiet NVs. 
Subsequent rule-based merging of short silences did not adequately resolve this issue. 
We also attempted to use the Montreal Forced Aligner (MFA) \cite{mcauliffe2017montreal} to align the audio with the scripts. However, due to the extended length of the recordings, MFA frequently failed to produce a successful alignment.

Ultimately, we developed a novel solution by leveraging an audio-capable LLM \cite{comanici2025gemini25pushingfrontier} to generate sentence-level timestamps for the audio files, which proved to be highly effective. Based on these timestamps, we segmented the recordings and then validated each sample using an ASR model \cite{gao2022paraformer}. Only samples with a Character Error Rate (CER) below a predefined threshold were retained. A key insight from this process is that for tasks requiring semantic-level segmentation, purely acoustic-based tools like MFA or VAD can be inadequate; a semantically-aware approach is often necessary.

\begin{figure}[t]
\centering
\includegraphics[width=\columnwidth]{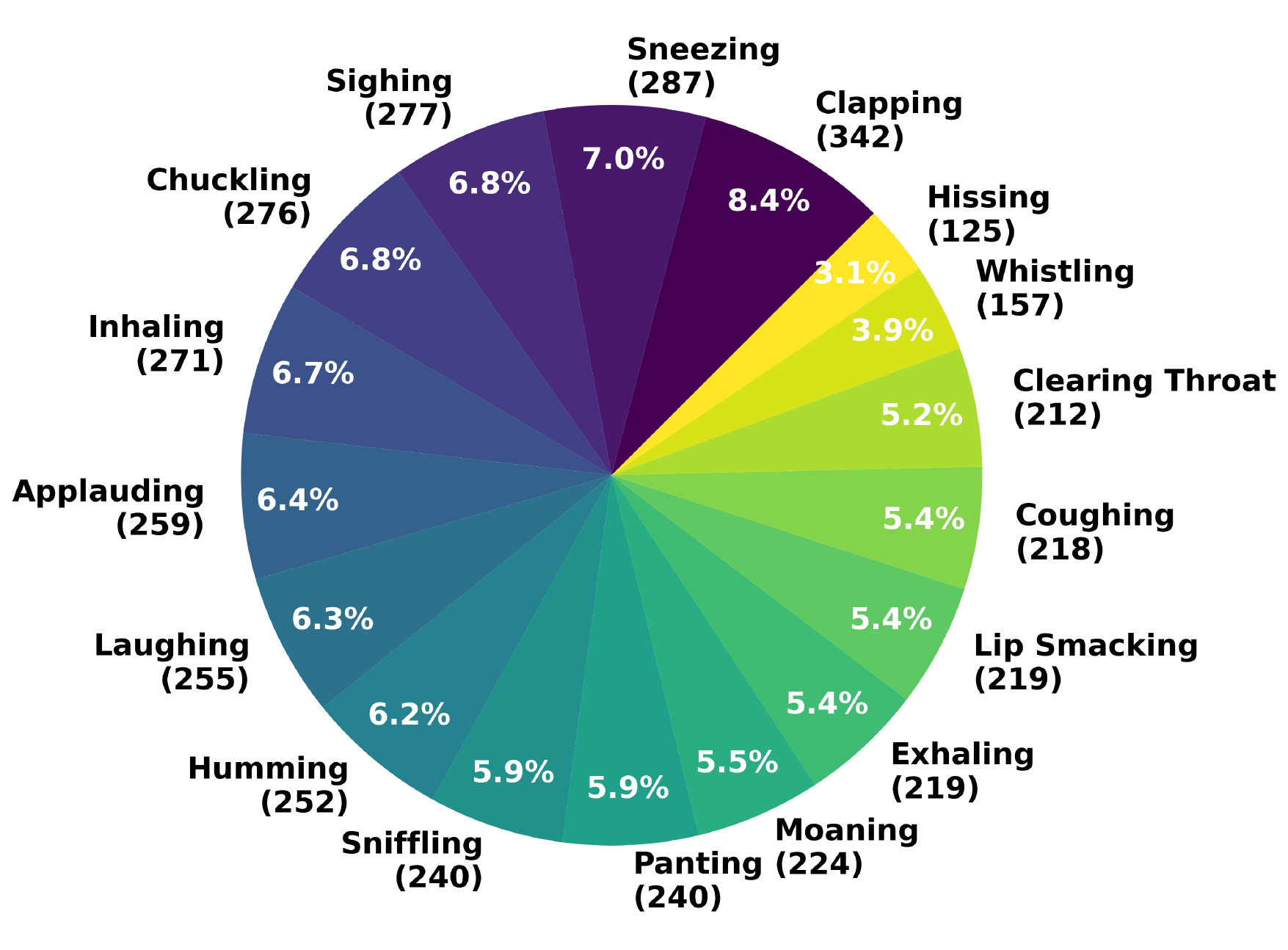} 
\caption{Distribution of the 17 NV categories in the MNV-17 dataset.}
\label{fig:nv_distribution}
\end{figure}

After post-processing, the dataset consists of 2,444 samples. The distribution of samples containing one, two, or three NVs is 1,272 (52.0\%), 715 (29.3\%), and 457 (18.7\%), respectively. The detailed distribution across individual NV categories is illustrated in Figure \ref{fig:nv_distribution}.

Class imbalance poses a significant risk of introducing model bias, as models may become overly sensitive to majority classes while neglecting less frequent ones \cite{mai2025aa}. To evaluate the balance of our dataset, we use the ratio of the most frequent class to the least frequent class (Max/Min Ratio), where a lower ratio indicates better balance. As shown in Table \ref{tab:dataset_comparison}, MNV-17 achieves the lowest ratio compared to other existing corpora, underscoring its superior class balance.

The final dataset is organized into 49 directories, one for each speaker. To ensure that our models learn the general acoustic patterns of NVs rather than overfitting to the vocal habits of specific individuals, we implemented a speaker-independent split for the training, validation, and test sets. We recommend that the male and female participants with IDs '00' and '01' (4 speakers total) be used for the validation set, and those with IDs '02' and '03' (4 speakers total) be used for the test set. The remaining 41 speakers constitute the training set. All subsequent experiments reported in this paper adhere to this data-splitting protocol.

\section{Experiments}
This section evaluates the performance of four mainstream ASR architectures on the MNV-17 dataset from three perspectives.

\subsection{Joint ASR and NV Recognition}
This experiment jointly evaluates both the ASR and NV recognition capabilities of the models. We finetuned four pretrained systems on the MNV-17 training set: SenseVoice \cite{an2024funaudiollm}, Paraformer (zh version), Qwen2-Audio (instruct version) \cite{chu2024qwen2}, and Qwen2.5-omni \cite{xu2025qwen2}. For each model, the checkpoint with the best performance on the validation set was selected for inference on the test set. Following standard practice \cite{radford2023robust, baevski2020wav2vec}, punctuation and spaces were removed from the transcripts before calculating the Character Error Rate (CER). Each NV label was treated as a single character, and the final CER was averaged across all samples in the test set.

For the autoregressive models, Qwen2-Audio and Qwen2.5-Omni, we employed the parameter-efficient LoRA method for fine-tuning. Both models were trained for 3 epochs using an Adam optimizer with a learning rate of 1e-4. The LoRA rank was set to 8 and alpha to 32, targeting all linear modules. To ensure stable training, we used gradient accumulation, resulting in an effective batch size of 64 for Qwen2-Audio and 32 for Qwen2.5-Omni.

For the non-autoregressive models, Paraformer and SenseVoice, we conducted full-parameter fine-tuning for a maximum of 50 epochs. These models were trained using an Adam optimizer with a learning rate of 2e-4. A warmup scheduler was applied for the first 30,000 steps to stabilize the training process. Dynamic token-based batching was utilized, with each batch containing approximately 20,000 tokens.

The results are presented in Table \ref{tab:joint_performance}.

\begin{table}[h]
\centering
\caption{Joint ASR and NV recognition performance. CER is calculated with NV labels treated as single characters.}
\label{tab:joint_performance}
\begin{tabular}{lc}
\toprule
\textbf{Model} & \textbf{CER (\%)} \\
\midrule
SenseVoice & 8.71 \\
Paraformer & 5.70 \\
Qwen2-Audio & 4.84 \\
Qwen2.5-omni & \textbf{3.60} \\
\bottomrule
\end{tabular}
\end{table}

The results reveal a performance difference that can be attributed to the models' architectural designs. The autoregressive Large Audio Models, Qwen2-Audio and Qwen2.5-omni, significantly outperform their non-autoregressive counterparts, SenseVoice and Paraformer. This advantage may arise from the inherent flexibility of the autoregressive framework. By generating the output sequence token by token, these models can more effectively integrate the recognition of non-lexical NV events alongside standard lexical transcription, as each new prediction is conditioned on the rich, mixed context of previously generated words and NV tags. In contrast, non-autoregressive models, which are optimized for efficient, parallelized prediction, might be less effective at capturing the complex, sequential interaction between semantic content and discrete nonverbal events within a unified output. The leading performance of Qwen2.5-omni, which achieved the lowest CER of 3.60\%, underscores the potential of large-scale autoregressive architectures for this complex joint task.

\subsection{NV Recognition Accuracy}
This subsection isolates and evaluates the models' ability to accurately identify NV events. Using the models finetuned in the previous experiment, we assessed their performance based on a strict accuracy metric. For each test sample, we extracted the type, quantity, and sequential order of all NV events from the ground truth. A prediction was considered correct only if it exactly matched the ground truth in all three aspects—type, count, and order.

The results are shown in Table \ref{tab:nv_accuracy}.

\begin{table}[h]
\centering
\caption{NV recognition accuracy. A prediction is correct only if the type, count, and order of NVs match the ground truth exactly.}
\label{tab:nv_accuracy}
\begin{tabular}{lc}
\toprule
\textbf{Model} & \textbf{Accuracy (\%)} \\
\midrule
Paraformer & 28.64 \\
Qwen2-Audio & 56.28 \\
SenseVoice & \textbf{57.29} \\
Qwen2.5-omni & \textbf{57.29} \\
\bottomrule
\end{tabular}
\end{table}

The performance disparity on NV recognition stems directly from the models' pretraining paradigms. Paraformer, exclusively pretrained for ASR, struggles significantly (28.64\% accuracy) as its objective was limited to lexical transcription.

In contrast, the top-performing models, SenseVoice, Qwen2-Audio, and Qwen2.5-omni, all were pretrained across diverse audio tasks. This foundational training equipped them with generalized acoustic representations and strong sensitivity to various acoustic events. Fine-tuning these robust models on MNV-17 effectively activated and refined their latent nonverbal recognition capabilities, demonstrating that our dataset excels at leveraging pre-existing multi-task knowledge for the specific challenge of expressive ASR.

\subsection{Impact of NV Integration on ASR Performance}
A critical question is whether adding NV recognition capabilities affects a model's core ASR performance. To investigate this, we conducted a comparative analysis. First, we trained baseline ASR models on the MNV-17 dataset with all NV labels removed from the transcripts. Second, we took the predictions from the NV-aware models (trained in Section 3.1), stripped the NV labels from their output, and calculated the CER against the NV-free ground truth. This approach allows for a fair and direct comparison of ASR performance with and without the NV recognition task.

The results of this comparison are detailed in Table \ref{tab:asr_impact}.
\begin{table}[h]
\centering
\caption{Comparison of CER (\%) for models trained with and without NV recognition capability. The CER for the NV-aware models was calculated after removing NV tags from their predictions.}
\label{tab:asr_impact}
\begin{tabular}{lcc}
\toprule
\textbf{Model} & \textbf{Non-NV Model} & \textbf{NV-aware Model} \\
\midrule
SenseVoice & 7.01 & 7.48 \\
Paraformer & 1.66 & 2.88 \\
Qwen2-Audio & 3.05 & 2.60 \\
Qwen2.5-omni & 1.53 & 1.72 \\
\bottomrule
\end{tabular}
\end{table}

The analysis reveals a significant decline in ASR performance for the model pretrained exclusively on ASR. Paraformer's transcription accuracy degraded significantly, with its CER increasing by 1.22\% when tasked with joint recognition.

In contrast, models with multi-task pre-training backgrounds demonstrated far greater resilience. SenseVoice and Qwen2.5-omni experienced only minor CER increases of 0.47\% and 0.19\%, respectively, indicating they can absorb the new capability with negligible impact. Most notably, the ASR performance of Qwen2-Audio actually improved, with its CER dropping from 3.05\% to 2.60\%.

For autoregressive Large Audio Models like Qwen2-Audio, this mutually beneficial effect likely occurs because a correctly identified NV is generated as a token in the output sequence. This token then becomes an explicit part of the context used to predict subsequent words, allowing the model to leverage the non-lexical cue to improve the transcription accuracy of the surrounding speech.
\section{Conclusion}

In this paper, we addressed the gap in mainstream ASR systems' ability to recognize nonverbal vocalizations (NVs) by introducing MNV-17, a 7.55-hour, high-quality performative Mandarin speech dataset. Featuring 17 distinct and well-balanced NV categories, MNV-17 provides a robust resource for developing more expressive ASR. Our experiments demonstrate that large, multi-task audio models are particularly adept at the joint task of NV recognition and speech transcription, with Qwen2.5-omni achieving a superior joint CER of 3.60\%. Notably, our findings indicate that integrating NV recognition capabilities does not degrade, and can even enhance, the core ASR performance of these advanced models. The dataset and pretrained model checkpoints are publicly available to facilitate future research in expressive ASR.

\section{Acknowledgments}

This work was supported by the Guangdong Basic and Applied Basic Research Foundation (2025A1515011203), the Guangdong Provincial Key Laboratory of Human Digital Twin (2022B1212010004), and the Nansha Key Project (2022ZD011).

\bibliographystyle{IEEEbib}
\bibliography{strings,main}

\end{document}